\documentclass{aa}

\newcommand{\vsi}{$v\,\mathrm{sin}\,i$}
\newcommand{\gdor}{$\gamma$~Dor}

\usepackage{graphicx}
\usepackage{txfonts}
\newcommand{\rev}[1]{#1}
\newcommand{\revb}[1]{#1}
\usepackage{siunitx}
\usepackage[varvw]{newtxmath}

\begin{document}

\title{Relation of Rotation and Pulsation Properties in $\gamma$~Doradus Stars}

\author{Á. Sódor,
        E. Plachy, \and
        Zs. Bognár
}

\institute{
  Konkoly Observatory, HUN-REN Research Centre for Astronomy and Earth Sciences, MTA Centre of Excellence, H-1121 Budapest, Konkoly Thege Mikl\'os \'ut 15-17, Hungary\\
  \email{sodor@konkoly.hu}
}

\date{Received August 15, 2026; accepted -- --, 2026}

\abstract
    {Connection between stellar rotation and pulsation amplitude has been proposed for several types of pulsating variable stars, but no such relation has been established for $\gamma$~Doradus stars from spectroscopic \vsi\ measurements. \rev{A high-amplitude $\gamma$~Doradus subclass was also suggested recently, which had not been explored by spectroscopic means before.}}
    {We investigate the connection between rotation and pulsational properties of $\gamma$~Doradus stars using a homogeneous spectroscopic sample with available TESS light curves.}
    {We obtained new stellar spectra with the ACE spectrograph on the 1-m RCC telescope of Konkoly Observatory and measured \vsi\ for a sample of 124 targets. For target selection, Gaia classifications were combined with TESS photometry. We used Fourier methods to determine the dominant frequencies from all available TESS light curves through sector 98. \rev{We also performed rotation- and inclination-dependent mode visibility modelling relying on GYRE Hough functions with TAR visibilities for the interpretation of our results.}
    }
    {We find a \revb{decreasing} upper envelope in both the peak-to-peak amplitude and the dominant-component semi-amplitude as a function of \vsi: high-\vsi\ stars show only low photometric amplitudes, whereas high-amplitude stars have low \vsi. We also confirm a positive relation between \vsi\ and the dominant pulsation frequency. The highest observed amplitudes \revb{also} decrease towards higher dominant frequencies. \rev{Visibility modelling indicates that these trends cannot be explained by geometric effects alone, suggesting that the intrinsic pulsation amplitudes of \gdor\ stars physically decrease with increasing equatorial rotational velocity. Furthermore, the strong suppression of zonal dipole mode visibility at high rotation suggests that high-amplitude \gdor\ stars are likely slow rotators or moderate rotators viewed nearly pole-on.}
    }
    {}


\maketitle
\nolinenumbers

\section{Introduction}

$\gamma$~Doradus (\gdor\ hereafter) stars are intermediate-mass, \revb{late-A} to F spectral-type, main-sequence pulsators whose observed variability is dominated by high-order gravity modes. \revb{Classically, their multiperiodic variability has periods of roughly 0.4\,--\,3 d (frequencies of about 0.3\,--\,2.5 d$^{-1}$) and optical amplitudes up to about 0.1 mag \citep{Kaye1999}.} Their multiperiodic light curves offer access to stellar structure and rotation by means of asteroseismology. Space photometry combined with spectroscopy has enabled homogeneous studies of substantial samples \citep{Uytterhoeven2011,Tkachenko2013}. Rotation is consequently central to the interpretation of both the frequencies and the observed photometric properties of these stars.

Rotation modifies gravity-mode pulsations through the Coriolis force and through the transformation between the co-rotating and inertial reference frames \citep{Li_2020_gmode_near-core-rot}. For stars with identified period-spacing patterns, modelling under the traditional approximation can constrain near-core rotation. In such mode-identified samples, the dominant inertial-frame mode frequency increases with inferred near-core rotation \citep{VanReeth2016}. \textit{Kepler} and TESS period-spacing catalogues have extended this seismic context to hundreds of \gdor\ stars \citep{Li_2020_gmode_near-core-rot,Garcia2022InternalRotation}. These inferences, however, require mode-pattern information, which, in turn, requires sufficient frequency resolution, available only on annual-scale time series, due to the low pulsation frequencies. Shorter, month-scale data, however, permits only determining several of the highest-amplitude, dominant pulsation modes.

Spectroscopic \vsi\ provides a complementary observational constraint, but it is not a direct measurement of either equatorial or near-core rotation. It combines surface equatorial velocity, $v_\mathrm{eq}$, with the unknown inclination, $i$, and it does not trace the rotation of the mode cavity without additional assumptions. Consistent with the expected influence of rotation on observed frequencies, \citet{VanReeth2015} found a positive association between the dominant $g$-mode frequency and \vsi\ in a \textit{Kepler} sample selected for detectable period-spacing patterns. This result motivates an examination of frequency--rotation behaviour in other samples.

The interpretation of photometric pulsation amplitude is complicated. An observed light curve amplitude is not equivalent to an intrinsic mode amplitude or mode energy: it can depend on the selected amplitude metric, inclination-dependent geometric mode visibility, mode cancellation, frequency extraction, time baseline, and the interaction between multiple modes. Rotation-dependent angular eigenfunctions and disk integration can also alter the observed visibility of low-frequency modes \citep{Townsend2003LightVariations,Townsend2003AngularModes}. In particular, a peak-to-peak light-curve range and the amplitude of a dominant Fourier component quantify different aspects of the observed variability. Any comparison with rotation must therefore account for both projection in \vsi\ and observational or geometric effects on the measured amplitude.

\rev{\citet{Paunzen2020HAGDOR} proposed a new subclass of the high-amplitude \gdor\ (HAGDOR) stars among \gdor\ pulsators, defined as objects by a Johnson-$V$ total peak-to-peak range above 0.1 mag, and showed that superposed base, combination, and harmonic frequencies can produce large total ranges even when the largest primary-frequency component remains below that threshold. These reports provide a further motivation to examine unusually strong observed variability, testing whether the observed amplitude range, including the highest amplitudes, is related to the projected stellar rotation measured by \vsi.}

\rev{Testing for an observed amplitude--\vsi\ relation is useful even though neither quantity alone gives the intrinsic pulsation amplitude or the true equatorial rotation rate. Such an association would show that rotation, inclination-dependent projection, and mode visibility may jointly shape the population that is detected and classified as \gdor\ variables. This matters because photometric amplitudes influence target detectability and amplitude-selected samples, whereas \vsi\ is an inclination-dependent surface-rotation observable. Conversely, the absence of a simple relation would constrain overly simplified one-mode or geometric explanations. Establishing the observational behaviour, together with its selection and visibility limitations, is therefore necessary before attributing any amplitude trend to intrinsic pulsation physics.}

A recent study by \citet{Wang2025} investigated the relation between photometric variability and \vsi\ for a sample of 367 \gdor\ stars using TESS photometry and LAMOST spectroscopy. They found no clear or statistically significant relation between the TESS-band peak-to-peak amplitude associated with the primary period and \vsi. However, further investigation using independent samples, spectroscopic measurements, and approaches to characterising photometric amplitudes may provide additional insight into this relation.

In this work, we use our own spectroscopic \vsi\ measurements for 124 \gdor\ candidates, combined with TESS photometry (\citealt{Ricker2015TESS}) to investigate the relation between the observed rotation and pulsation properties.

\section{Data and Target Selection}
\label{sect:data}

We analyse photometric and spectroscopic pulsational and rotational properties of a sample of \gdor\ pulsators in the present study. Our targets were selected from a larger sample of \gdor\ stars, compiled as part of a separate project (Plachy et al., in prep.). Here, we briefly describe the steps relevant to the selection and characterisation of the presently analysed sample.

\subsection{Photometry from TESS}

The initial target list was based on the common Gaia DR3 variability classification flag of $\delta$ Scuti, \gdor, and SX Phoenicis variables, \texttt{DSCT\textbar GDOR\textbar SXPHE} \citep{Gaia2023_DR3}, which contains approximately 748 thousand sources \citep{Eyer_2023,Rimoldini_2023}. To identify \gdor\ pulsators among these objects and characterise their light variations, we used TESS full-frame image photometry. We obtained \revb{Simple} Aperture Photometry (SAP) light curves from the MIT Quick Look Pipeline (QLP; \citealt{qlp1-2020,qlp2-2021,qlp3-2022}) for targets brighter than $T_{\rm mag}=13.5$. TESS QLP light curves were available for more than 300 thousand sources from the initial Gaia sample.

For each light curve, we calculated Fourier components with the LCfit program developed by \cite{Sodor-2012}, adopting a signal-to-noise threshold of S/N $>4$. Candidate \gdor\ variables were required to have a dominant frequency between 0.3 and 3 d$^{-1}$, a dominant Fourier amplitude greater than 0.005 in normalized flux, and an amplitude ratio $A_2/A_1>0.1$ between the two strongest frequency components. The latter criterion ensured that the selected pulsators showed clear multiperiodic variability.

As part of the cleaning process, we rejected potential eclipsing binaries for which the frequency ratio of the two strongest peaks, $f_1/f_2$, was consistent with either 2 or 0.5 within the Rayleigh frequency resolution. To reduce the effects of blending and contamination on the measured amplitudes, we also excluded targets with a brighter neighbour within 2 arcsec. The remaining candidates were visually inspected, and misclassified objects, including eclipsing binaries, rotational variables, and apparently single-mode pulsators, were removed, together with light curves affected by instrumental trends, strong noise, or numerous outliers. Finally, we excluded hybrid \gdor--$\delta$ Scuti stars showing significant high-frequency $\delta$ Scuti variability and \gdor\ stars exhibiting eclipses.

From the resulting sample of approximately 3600 \gdor\ stars, we selected 125 targets for spectroscopic observations at Piszkéstető Observatory, taking into account their observability and aiming to cover a wide range of pulsation amplitudes. These stars constitute the sample analysed in this study.

In our analysis, we use the following parameters derived from the TESS light curves up to sector 98:
\begin{itemize}
    \item Dominant and second dominant amplitude, $A_1$ and $A_2$, respectively: the semi-amplitudes of the two largest-amplitude harmonic components of the light variations. They are unitless, expressed in flux relative to the mean flux of the variable star.
    \item Dominant frequency, $f_1$: the frequency of the dominant harmonic component, expressed in cycles per day (d$^{-1}$).
    \item Peak-to-peak amplitude (PTP): the difference between the highest and lowest observed relative flux of the target.
\end{itemize}

We use both PTP and $A_1$ because they characterise different aspects of the observed light variation. PTP measures the total observed flux range and is affected by the combined contribution, relative phases, \revb{and visibilities} of the pulsation modes, \revb{combination and harmonic frequencies}, whereas $A_1$ measures the \revb{visible amplitude} of the single dominant component. Both quantities can depend on inclination and geometric mode cancellation, but not necessarily in the same way. Therefore, a relation that is present for both metrics is less likely to be an artefact of a particular amplitude definition.

Quoted and plotted uncertainties are the $1\sigma$ standard deviations of the respective parameters. Note that the upward uncertainty of PTP is hard to quantify, as the sector sampling of TESS data is both limited and \revb{highly diverse} for the targets. There is always some chance that a higher PTP could be observed in a later sector, or would have been observed in a previous, not covered sector, and this chance is larger for stars observed in fewer sectors. Therefore, one has to keep in mind that this metric is only a lower estimate in this sense.

\subsection{Spectroscopic Target Selection and Observations}

\begin{table*}[!t]
\centering
\caption{Example rows from the spectroscopic target catalogue, listing spectroscopic and photometric parameters, based on their Konkoly ACE/RCC spectroscopy and TESS QLP SAP photometry. The full table is available electronically only, at the CDS via the VizieR service. $T_{\rm mag}$ is the TESS magnitude. $f_1$ and $A_1$ are the frequency and Fourier semi-amplitude of the dominant pulsation component extracted from the combined sectors of the TESS light curve. PTP is the peak-to-peak light variation in relative flux. Errors in parentheses are in the units of the least significant decimal digits of the corresponding measurements.
}\label{tab:spec124catalogue-sample}
\begin{tabular}{r S[table-format=1.2] S[table-format=1.6(4)] S[table-format=1.5(3)] S[table-format=1.5(3)] S[table-format=3.1(2)] c p{0.23\linewidth}}
\hline\hline
TIC ID & {$T_{\rm mag}$} & {$f_1$} & {$A_1$} & {PTP} & {$v\sin i$} & $N_{\rm sec}$ & Notes \\
{} & [mag] & {[d$^{-1}$]} & {[relative flux]} & {[relative flux]} & {[km s$^{-1}$]} & {} & {} \\
\hline
9375080 & 7.93 & 0.542281(19) & 0.01266(23) & 0.06450(28) & 17.7(5) & 2 & -- \\
10290108 & 9.79 & 0.906160(4) & 0.01750(15) & 0.13640(170) & 15.2(27) & 5 & -- \\
15422557 & 9.00 & 1.059949(5) & 0.01190(13) & 0.10720(120) & 10.9(28) & 3 & HGDS; ZhGD \\
18114983 & 9.85 & 2.266059(3) & 0.00542(3) & 0.03370(73) & 143.7(63) & 4 & -- \\
20528414 & 10.88 & 0.609380(2) & 0.01931(11) & 0.12982(73) & 15.6(53) & 5 & -- \\
21146971 & 9.68 & 2.160886(2) & 0.00543(2) & 0.03330(150) & 87.6(39) & 5 & -- \\
22635653 & 9.63 & 2.091695(5) & 0.00522(4) & 0.02540(68) & 64.3(34) & 3 & -- \\
22816037 & 8.07 & 2.226178(4) & 0.00715(6) & 0.08197(80) & 97.0(5) & 6 & NSS; RUWE = 11.75; ZhGD \\
23119777 & 9.83 & 0.404468(4) & 0.00679(4) & 0.03701(79) & 14.0(5) & 3 & NSS \\
24059469 & 9.24 & 1.639609(4) & 0.01124(9) & 0.09470(190) & 60.1(12) & 4 & -- \\
\multicolumn{1}{c}{\ldots} & \multicolumn{1}{c}{\ldots} & \multicolumn{1}{c}{\ldots} & \multicolumn{1}{c}{\ldots} & \multicolumn{1}{c}{\ldots} & \multicolumn{1}{c}{\ldots} & \multicolumn{1}{c}{\ldots} & \multicolumn{1}{l}{\ldots} \\
\hline
\end{tabular}
\begin{flushleft}\footnotesize
\textit{Notes.} NSS marks a Gaia DR3 non-single-star flag. RUWE is listed when $>1.4$. $N_{\rm sec}$ is the number of available TESS sector light curves. HGDS means Hey: GDOR/SPB -- target was classified by \cite{Hey_2024} as GDOR/SPB; ZhGD means Zhou: GDOR -- target was classified by \cite{Zhou_2024} as GDOR. This is a compacted version of the full electronic table with only columns relevant for the analysis.
\end{flushleft}
\end{table*}

We considered bright targets with $T_{\rm mag}<11.0$ and DEC $>-10^\circ$ for spectroscopic observations, as they are preferable and available for our instruments. For a pilot study, we initially selected two sub-samples from the low and high ends of the PTP amplitude range ($0.02<\mathrm{PTP}_\mathrm{low}<0.05$ and $0.1<\mathrm{PTP}_\mathrm{high}$), among the brightest targets observable at that particular season. Later, we extended the sample with targets from other seasons, fainter objects and targets filling in the initial PTP amplitude gap. However, due to this target selection strategy, the amplitude distribution of our spectroscopic sample is not representative for the whole photometric sample, but covers the broadest available PTP amplitude range. Altogether, 125 targets were successfully observed.

We obtained stellar spectra of the selected targets with the fibre-fed ACE spectrograph mounted on the 1-m RCC telescope of Konkoly Observatory operating at Piszkés-tető Mountain Station, working in the wavelength range of 4150\,--\,9150\,Å, with a
resolution of 20\,000. A thorium–argon (Th–Ar) lamp was used for wavelength calibration between each target acquisition. See further details on the instrument, its capabilities and the basic spectroscopic data processing in \cite{Derekas2017_v1154Cyg}.

We calculated average line-profiles in radial velocity (RV) space from the metallic lines by cross-correlating each individual observed spectrum with a synthetic, non-broadened stellar spectrum of F0 spectral type, solar metallicity, calculated by Synspec v49 \citep{HubenyLanz1995_SynSpec}. This resulted in a weighted average line profile, where the weights were the equivalent widths from the synthetic spectrum. Due to blended lines and spectral normalization irregularities, we re-normalized the cross-correlated line profiles before further processing.

Finally, $v_\mathrm{rad}$ and \vsi\ were determined using a simple Python script that fitted the line profiles, taking into account rotational line broadening, micro- and macroturbulence, relying on the \texttt{tools} package of Phoebe 2.0-alpha \citep{Prsa2016_Phoebe2.0}. However, the precision of our instrument does not permit an accurate micro- and macroturbulence deduction, and these values and \vsi\ are also largely degenerate. We found that $v_\mathrm{rad}$ is virtually insensitive, and \vsi\ is only slightly sensitive to the turbulences. Therefore, for each individual observation, we performed the line-profile fits with four combinations of two fixed microturbulence values of 1\,km/s and 2\,km/s, and two fixed macroturbulence values of 3\,km/s and 10\,km/s, and incorporated the resulting uncertainties in the total error budget. \rev{The adopted 1\,--\,2\,km/s microturbulence bracket was motivated by previous \gdor\ spectral analyses that fixed 1.5\,km/s for rotationally blended spectra and 2\,km/s where the available spectra could not constrain microturbulence independently \citep{Bruntt2008,Tkachenko2013}. A directly applicable macroturbulence calibration is unavailable for \gdor\ dwarfs, and unresolved non-rotational broadening can be partly absorbed by a fitted \vsi\ in late-A/F spectra \citep{Landstreet2009Macroturbulence}. We therefore use the 3 and 10\,km/s macroturbulence values solely as a deliberately broad line-profile sensitivity bracket. Based on these different fixed values, in the total \vsi\ error budget, the contributions of micro- and macroturbulence are generally 5\% and 12\%, respectively, when measured relative to the variation among repeated observations of the same target.}

Note that we provide RV measurements here, but we do not investigate them and their possible variations further, as the observing strategy was not designed to detect binaries based on RV variability.

At the same time, visual inspection of the line profiles indicated double-lined spectroscopic binary (SB2) nature of only one target -- TIC\,307989264. We therefore excluded this from the analysis. The final sample comprises 124 \gdor\ stars. We refer to this set as spectroscopic sample. The log of each individual observation, altogether 309 exposures, along with the exposure times and RVs are summarized in an electronic supplementary table. Table~\ref{tab:obslog-sample} demonstrates a sample of the full electronic table. The spectroscopic observations were acquired between 17. March and 18 September 2025. Altogether 309 individual exposures were collected on 20 observing nights.

\begin{table*}
\centering
\caption{Example rows from the spectroscopic observation log, along with the individual observed radial velocities (RV). The full table is available electronically only, at the CDS via the VizieR service. We retained 309 individual observations on 125 targets, including TIC\,307989264, which was excluded from the final spectroscopic \vsi\ sample due to its clear SB2 nature. Each row corresponds to a single observation/exposure. HJD is the Heliocentric Julian Date corresponding to the mid-exposure. The RV uncertainties are indicated in parentheses in the unit of the least significant digit (0.1\,km s$^{-1}$).}
\label{tab:obslog-sample}
\begin{tabular}{c r S[table-format=2.1] S[table-format=3.1(2)]}

\hline\hline
HJD & TIC ID & {Exposure time} & {RV} \\
{[d]} & {} & {[min]} & {[km s$^{-1}$]} \\
\hline
2460752.30541 & 426669407 & 20.0 & -25.5(12) \\
2460752.31936 & 426669407 & 20.0 & -26.5(5) \\
2460752.33330 & 426669407 & 20.0 & -29.6(5) \\
2460752.34823 & 26045729 & 5.0 & -6.7(13) \\
2460752.35174 & 26045729 & 5.0 & -9.2(10) \\
2460752.35525 & 26045729 & 5.0 & -8.6(11) \\
2460752.35885 & 26045729 & 5.0 & -12.5(10) \\
2460752.36236 & 26045729 & 5.0 & -8.5(10) \\
2460752.36587 & 26045729 & 5.0 & -10.9(13) \\
2460752.38212 & 69370154 & 20.0 & 43.2(6) \\
\multicolumn{1}{c}{\ldots} & \multicolumn{1}{c}{\ldots} & \multicolumn{1}{c}{\ldots} & \multicolumn{1}{c}{\ldots} \\
\hline
\end{tabular}
\end{table*}

Although Gaia DR3 flags 12 of our spectroscopic targets as multiple (NSS: Non-Single Star flag), and another 20 have $\mathrm{RUWE}>1.4$ value in the same catalogue, suggesting non-single or blended sources, their distribution does not show deviation from the rest of the sample in any of the investigated parameter combinations. Therefore, we flagged but included these 32 objects (26\,\%) in our sample on the basis that any significant light contamination or line-profile distortion from a possible binary companion or background star would probably have been apparent in the line profiles in the first place.

We also note here that we checked the SPOC CROWDSAP \citep{Jenkins2016_SPOC} crowding or light contamination indices for our sample, which was available on 122 targets, and found that only 5 have a significant flux contamination between 18 and 45\,\% (CROWDSAP between 0.55 and 0.82), while the median CROWDSAP value is 0.988, corresponding to 1.1\,\% relative flux contamination only. This can be attributed to the relatively high brightness of our sample, and to the selection criteria.
This, together with the above binarity check indicates that any light contamination source has only marginal effect on the observed pulsation amplitudes, expressed in relative flux compared to the mean brightness of the targets.

The collected spectroscopic and photometric data on the spectroscopic sample is summarized in an electronic supplementary table. Table~\ref{tab:spec124catalogue-sample} demonstrates a sample of the full electronic table.

We cross-checked our spectroscopic targets against the published classifications of \cite{Hey_2024} and \cite{Zhou_2024}. We list all the classifications of these authors for the matching targets in Table~\ref{tab:spec124catalogue-sample}. With three exceptions, all matches were classified as \gdor\ (by \citealt{Zhou_2024}) or \gdor--SPB (by \citealt{Hey_2024}).

\section{Results}

Below we present our observational findings, as relations between the observed spectroscopic and photometric parameters.

\subsection{Observed Amplitude--Projected-Rotation Distributions}

\begin{figure*}[ht]
  \centering
  \includegraphics[width=0.97\columnwidth]{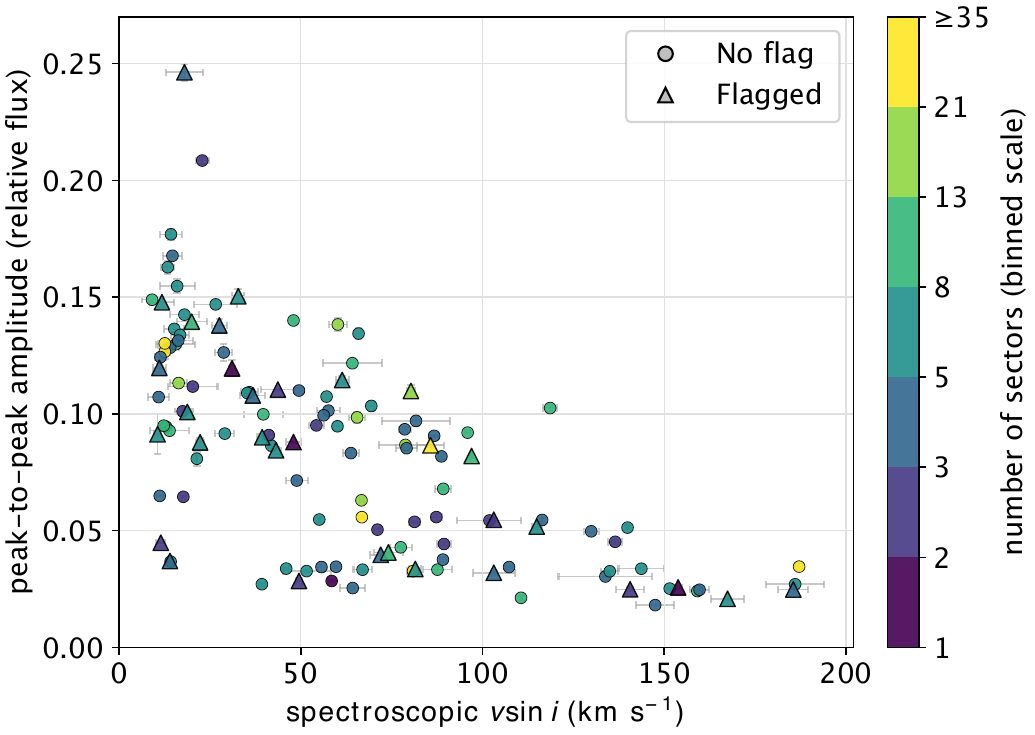}\hspace{4mm}
  \includegraphics[width=0.97\columnwidth]{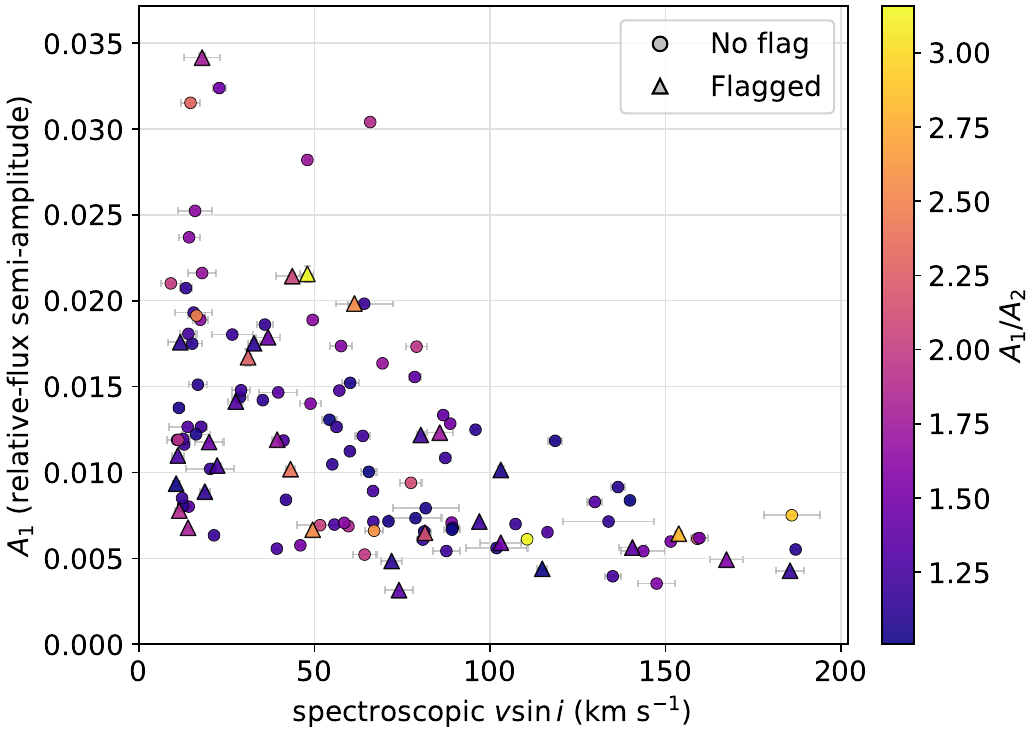}
  \caption{Amplitudes versus spectroscopic projected rotation velocity, \vsi, for the spectroscopic \gdor\ sample. Left-hand panel: PTP vs. \vsi, right-hand panel: $A_1$ vs. \vsi. ``Flagged'' targets have a Gaia NSS flag and/or RUWE $>1.4$, indicating possible flux and spectroscopic contamination. Colour codes the binned number of available, accepted TESS sectors, and $A_1/A_2$ amplitude dominance in the left- and right-hand panels, respectively.}
  \label{fig:amp-vsini-flagged}
\end{figure*}

Our main results are shown in the two panels of Fig.~\ref{fig:amp-vsini-flagged}, where the peak-to-peak, PTP, and the dominant, $A_1$, amplitudes are plotted against the \vsi\ measurements for the spectroscopic sample of \gdor\ variables. Both plots clearly demonstrate that high-\vsi, fast-rotating \gdor\ stars have low amplitudes, while low-\vsi\ \revb{objects}, that is, either slow rotators or fast rotators seen near pole-on, can have any amplitude from the lowest to the highest. This is true both for the total light variation and for the dominant mode amplitude. The same upper-envelope behaviour is present for both PTP and $A_1$. Since these metrics respond differently to the multiperiodic mode content and its geometric visibility, their qualitative agreement supports the robustness of the observed amplitude--\vsi\ distribution.

From the \revb{left-hand} panel of Fig.~\ref{fig:amp-vsini-flagged} we can see that the total PTP pulsational light variation of the fastest rotators, with \vsi\ $> 150\,\mathrm{km/s}$, does not exceed 0.035 relative flux in the TESS photometric band. At the same time, all pulsators exceeding 0.15 PTP amplitude in relative flux show \vsi\ below 25 km/s.

The 32 \revb{objects with potential} light contamination (binary or blended) are marked with triangles in the two panels of Fig.~\ref{fig:amp-vsini-flagged}, and are labelled as ``Flagged''; however, the distribution of these objects does not visibly deviate from the rest. Colour codes the binned number of available, accepted TESS sectors, and $A_1/A_2$ amplitude dominance in the left- and right-hand panels, respectively. The number of TESS sectors suggests the possible degree of underestimating the true largest possible PTP for any target. $A_1/A_2$ amplitude dominance is a rough indicator of how much of the pulsation energy is concentrated in the dominant mode. There is no apparent trend with any of these two metrics within our sample.

\subsection{Dominant Frequency and Projected Rotation}

\begin{figure}
   \centering
   \includegraphics[width=0.97\columnwidth]{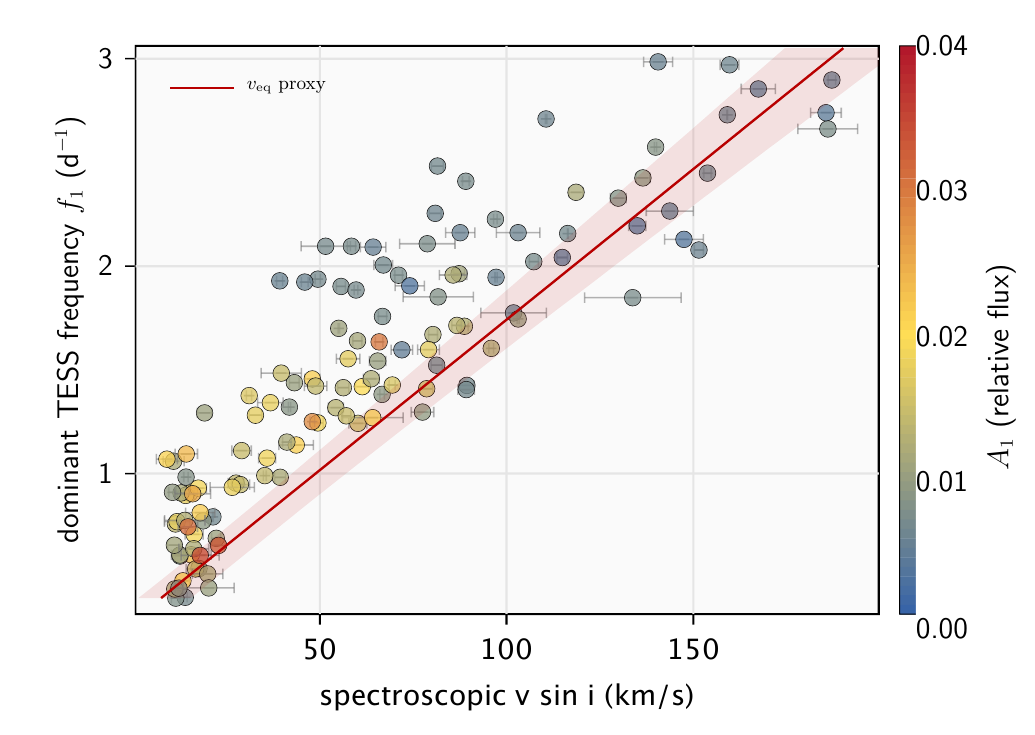}
   \caption{Dominant frequency, $f_1$, versus spectroscopic projected rotation velocity, \vsi, for the spectroscopic \gdor\ sample. Colour \revb{indicates} the dominant pulsation amplitude, $A_1$. \rev{Continuous red line and band denotes the estimated relation between the equatorial rotation speed ($v_\mathrm{eq}$) and the dominant frequency, and its 1-$\sigma$ uncertainty based on proxies, respectively (see discussion in Sect.~\ref{sect:discussion-qualitative}). Note that $v_\mathrm{eq}$ is represented on the same velocity scale as \vsi.}}
   \label{fig:f1-vs-vsini}
\end{figure}

Another important observation is presented in Fig.~\ref{fig:f1-vs-vsini}, showing the relation between the observed \vsi\ and the dominant frequency, $f_1$ for the spectroscopic sample. Here, both the bottom-right and top-left corners are unpopulated, and the two \revb{attributes} clearly correlate. Furthermore, Fig.~\ref{fig:f1-vs-vsini} also shows correlation of the dominant amplitude, $A_1$ with $f_1$, but this will be better visible in the next figure.

The fact that there are no low-frequency, high-\vsi\ targets in the sample, and a clear boundary can be drawn in the figure, with no data points of \gdor\ stars situated to its right, is easier to interpret. This means that the higher the dominant frequency is, the higher \vsi\ can be. As the inclination, $i$, is unknown, the true equatorial velocity, $v_\mathrm{eq}$ represents only an upper bound to \vsi, for each individual target; therefore, a data point can only fall to the left of this hypothetical boundary. However, our data is insufficient to give a mathematical formula for this limit, therefore the boundary remains qualitative.

\revb{Additionally}, the top-left of the diagram is also unpopulated in Fig.~\ref{fig:f1-vs-vsini}, that is, there is also no \gdor\ star with high dominant frequency \revb{and} low \vsi\ in our sample. This might be somewhat surprising, given \revb{that the unknown inclination could shift \vsi\ towards zero by an arbitrary amount}. To explain this observation, mode visibilities also have to be involved. We return to this problem in the Discussion, in Sect.~\ref{sect:discussion-qualitative}. This \revb{upper left} limit is also not \revb{well} defined in our sample \revb{and consequently} also remains qualitative.

\begin{figure*}[!th]
  \centering
  \includegraphics[width=0.97\columnwidth]{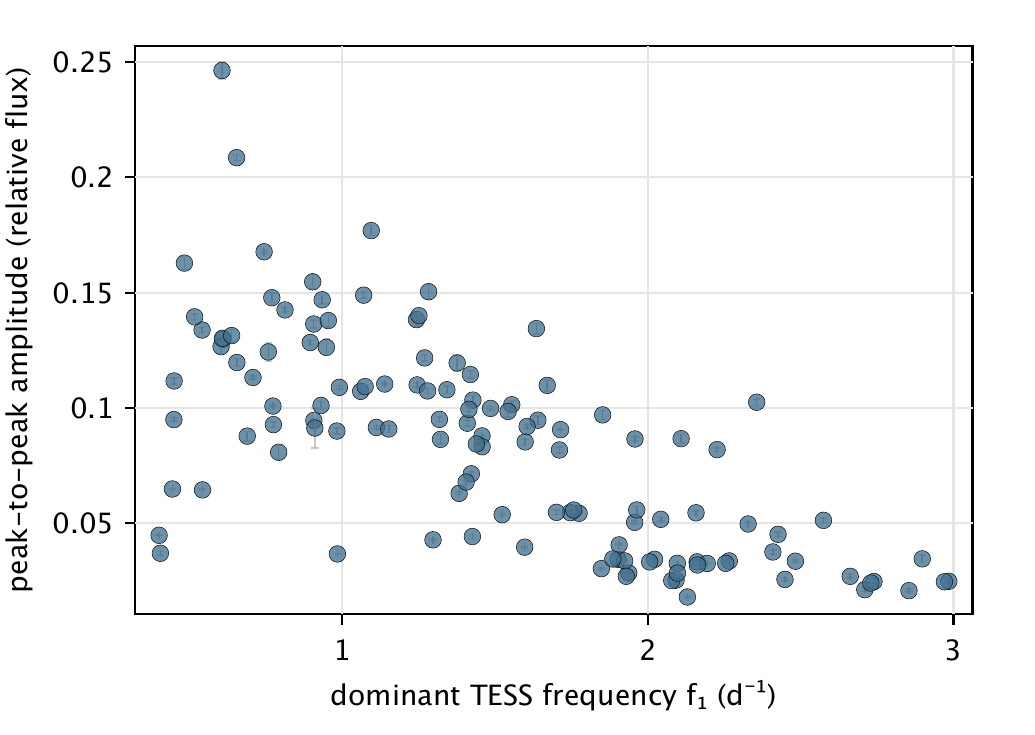}\hspace{4mm}
  \includegraphics[width=0.97\columnwidth]{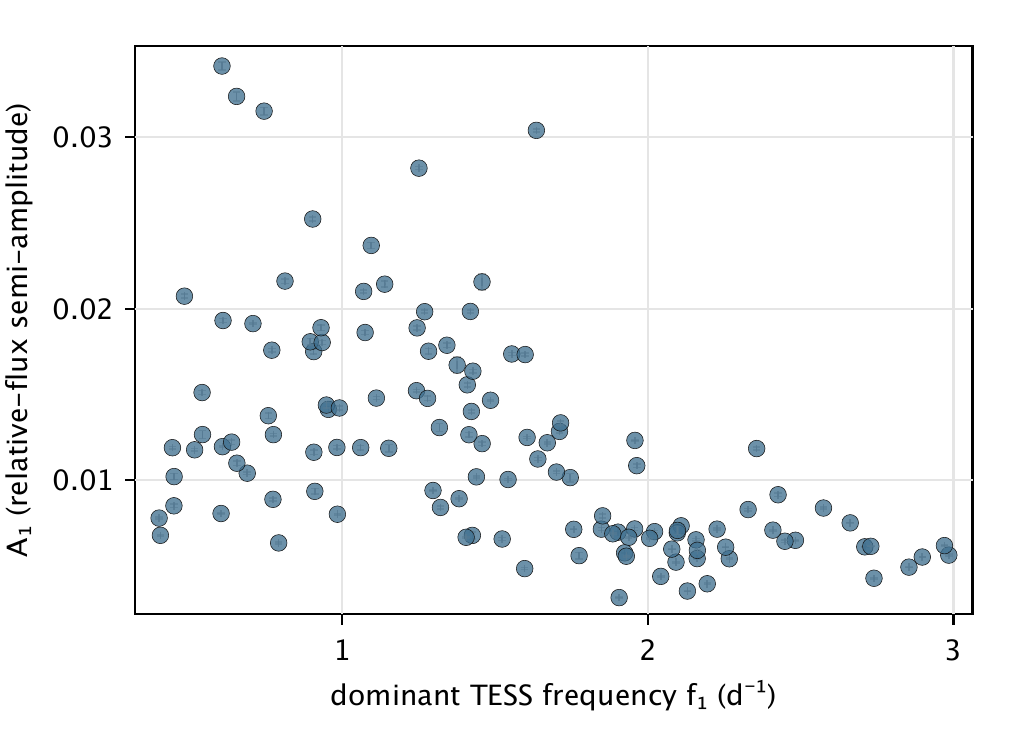}
  \caption{Peak-to-peak amplitude, PTP, and dominant-component semi-amplitude, $A_1$, versus dominant frequency, $f_1$, for the spectroscopic \gdor\ sample. Uncertainties are plotted, but in most cases are much smaller then the dot size.}
   \label{fig:amp_vs_f1}
\end{figure*}

\subsection{Dominant Frequency and Amplitude}

Another important relation between the pulsation amplitudes and frequencies is shown in the two panels of Fig.~\ref{fig:amp_vs_f1}. The top-right parts of the diagrams are completely unpopulated in both diagrams, and there is an upper-envelope-like boundary relation. The higher the dominant frequency is, the lower the corresponding amplitude limit gets, both for PTP and for $A_1$.

\section{Discussion}

\rev{We compare below our results to earlier publications, and discuss the possible interpretations of our findings. To investigate the geometric effects that may have influence on the observed relations, we developed a visibility toy model.}

\subsection{Comparison with Earlier Published Results}

The most direct comparison can be made with the published results of \cite{Wang2025}, who investigated a \rev{large} \gdor\ sample with TESS photometry and LAMOST spectroscopy available. Surprisingly, looking at their Fig.~7b, and according to their text, the \cite{Wang2025} \gdor\ sample does not show any meaningful relation between \vsi\ and $A_1$. Also, their \vsi\ range is significantly broader than ours, but the overlapping domain still does not show any obvious common feature. \rev{To explore this discrepancy, we looked into the results and methods of \cite{Wang2025} more thoroughly.}

\subsubsection{\rev{Comparison with \cite{Wang2025}}}

\rev{Table~1 of \cite{Wang2025} lists 367 "pure" \gdor\ stars -- according to their wording --, meaning no hybrids, only \gdor\ behaviour. We refer to these as Wang sample or Wang targets hereafter.}

\rev{First, we verified if these were indeed \gdor\ pulsators by reviewing their TESS light curves. Note that we were unable to faithfully reproduce the input light curve data of \cite{Wang2025}, as the authors did not indicate clearly the origin of their photometric data, including the pipeline and sectors used. Therefore, we chose to use all the available TESS QLP SAP light curves up to sector 98. We calculated periodograms for the combined sector light curves of each target, applied a low-pass filter to remove any variation below 0.2\,d$^{-1}$, and finally visually inspected them to verify classification. We excluded 98 targets, of which 22 are \gdor--$\delta$ Scuti hybrids, and the rest are different kinds of eclipsing binaries and other variables. The remaining 269 confirmed Wang \gdor\ targets were further subdivided into GDOR-A and B categories, representing certain and uncertain \gdor\ classification, respectively. The classification was based solely on visual inspection of the periodograms, relying on the number of visually distinguishable independent frequency peaks. The process is admittedly quick and somewhat subjective, although the exact details have no significant effect on the following findings. The results are detailed and the periodograms are shown in the supporting online material (see the data availability statement below).}

\begin{figure*}[!th]
\centering
  \includegraphics[width=0.97\columnwidth]{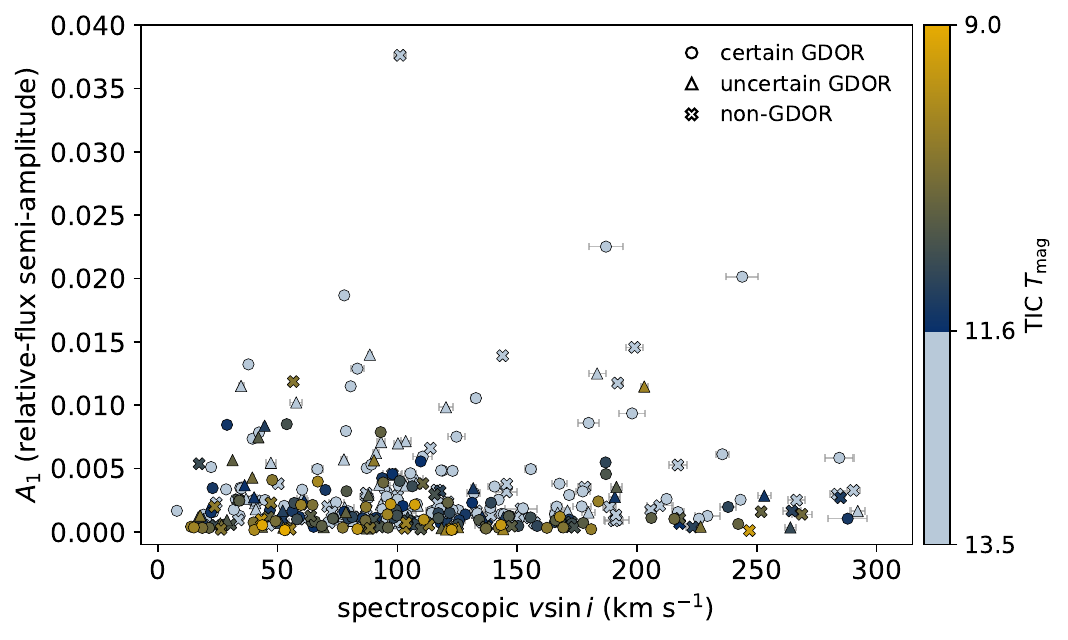}\hspace{4mm}
  \includegraphics[width=0.97\columnwidth]{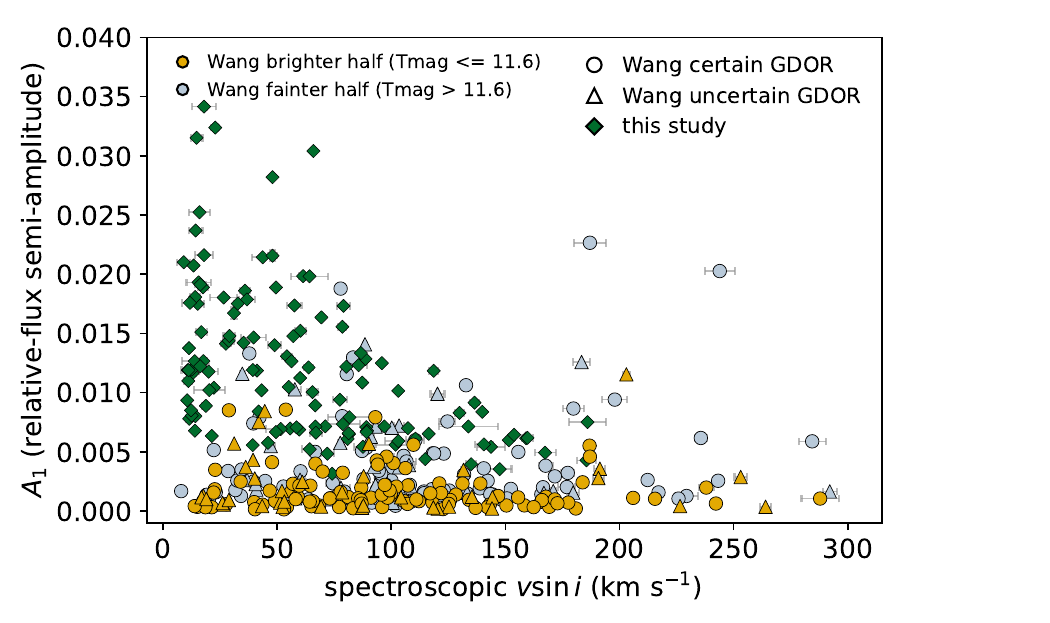}
  \caption{
  Investigation, reclassification and comparison of the \cite{Wang2025} 367-element "pure" \gdor\ sample to our spectroscopic sample of 124 targets. Left-hand panel: reclassification of the Wang targets. We distinguish certain and uncertain confirmed \gdor\ pulsators (circles and triangles, respectively) and identify 98 non-pure GDOR stars, including 22 hybrids (labelled as non-GDOR, cross symbols). Colour code indicates brightness, where the fainter half of the sample is plotted with grey symbols in both panels. Right-hand panel: only the 269 confirmed Wang GDORs (circles and triangles), where we distinguish the bright and faint halves of the sample with yellow and grey colours, respectively; together with our 124-element spectroscopic sample (green diamonds), on a tentative common amplitude scale.}
  \label{fig:wang-ace-context}
\end{figure*}

\rev{The $A_1$ vs. \vsi\ relation for the reclassified Wang targets is presented in the left-hand panel of Fig.~\ref{fig:wang-ace-context}. Here we plot the dominant amplitudes reported by \cite{Wang2025}, converted to a relative flux variation scale facilitating comparison with our sample. For the amplitude transformation, we calibrated a magnitude to relative flux function, based on 7 objects we found to be common with our larger TESS photometric sample of \gdor\ stars, mentioned in Sect.~\ref{sect:data} (Plachy et al., in prep.) Note that we have no objects in our 124-element spectroscopic sample common with the Wang targets. Compared with the original Fig. 7(b) of Wang et al. (2025), our figure not only uses converted amplitudes, but also includes revised classification and brightness information.}

\rev{We investigated the brightness distribution of the Wang targets and our sample, based on $T_{\rm mag}$ from the TESS Input Catalog (TIC-8; \citealt{Stassun2019TIC}). The brightness median and range of our spectroscopic sample are 8.71\,mag and 4.94\,--\,10.95\,mag, while for the Wang targets these are 11.59\,mag and 8.96\,--\,13.47\,mag. To assess the effect of brightness, we denote Wang targets fainter than their median by grey symbols. The left-hand panel of Fig.~\ref{fig:wang-ace-context} indicates that after reclassification, the brighter half of the Wang sample shows some resemblance to the general character of our $A_1$ vs. \vsi\ relation plotted in the right-hand panel of Fig.~\ref{fig:amp_vs_f1}, that is, an upper envelope decreasing with \vsi\ emerges. Two facts are, however, conspicuous.}

\rev{On the one hand, with a single exception, all the outliers towards the top right belong to the fainter half of the Wang sample. Similarly, among the 20 Wang targets above the 0.09 relative flux amplitude limit, only 2 are from the brighter half of the sample, which is 10\,\% instead of the expected 50\,\%. It is unclear whether these outliers in the fainter half of the sample arise from excessive uncertainties in the photometric or spectroscopic data. It might be the case that the uncertainties in the spectroscopic data of \cite{Wang2025} increase more rapidly towards fainter targets than their reported error estimates indicate. Alternatively, part of the uncertainty may also originate from the photometric data. We note that \cite{Wang2025} do not report uncertainties on the photometric amplitudes. We therefore cannot separate the contributions of photometric and spectroscopic uncertainties to the observed scatter, and we are also not in the position to attribute more realistic uncertainties to the Wang sample.}

\rev{On the other hand, considering only the brighter half of the Wang targets, these stars typically have much smaller pulsation semi-amplitudes than those in our sample. This discrepancy may arise from differences in target selection and in the TESS reduction pipelines used, including how they handle contamination.}

\rev{Although the above investigation points to several possible explanations for the discrepancy, ambiguities regarding the Wang sample and the performed data processing steps prevent us from assessing their individual contributions. We therefore do not pursue this reproducibility analysis further.}

\rev{The direct comparison of the $A_1$ vs. \vsi\ distribution of our 124-element spectroscopic sample and the 269 confirmed Wang \gdor\ targets is presented in the right-hand panel of Fig.~\ref{fig:wang-ace-context}. We note that the amplitudes are given only on a tentative common scale. This diagram shows that the brighter half of the Wang sample is fully consistent with our findings.}

\subsubsection{\rev{Comparison with Other Earlier Findings}}

\cite{Tkachenko2013} also investigated a sample of 69 \gdor\ pulsators, using spectroscopic \vsi, and \textit{Kepler} photometric  frequency and amplitude measurements. \cite{Tkachenko2013}[Fig.~5] corresponds to our Fig.~\ref{fig:f1-vs-vsini}, while their Fig.~6 corresponds to our Fig.~\ref{fig:amp_vs_f1}, left-hand panel. Both pairs of figures show qualitative agreement with our results. Although an upper envelope and unpopulated top-right corner is suggested by \cite{Tkachenko2013}[Fig.~6], they attempted to describe the dependence of the dominant amplitude on \vsi\ by linear regression, and concluded that there is no strong correlation between the two variables for \gdor\ stars.

Another asteroseismic study, performed by \rev{\cite{VanReeth2015}} also relied on \revb{practically the same sample, with two exceptions, that \cite{Tkachenko2013} published}. They also plotted dominant frequency vs. \vsi\ in the top panel of their Fig.~9, again, indicating qualitatively very similar results to our Fig~\ref{fig:f1-vs-vsini}.

\subsection{Qualitative Interpretation of the Relations Between Projected Rotation, Dominant Frequency and Amplitude}
\label{sect:discussion-qualitative}

Since \gdor\ stars pulsate in multiple non-radial modes, which all suffer different geometric cancellations based on their mode numbers and viewing angles, any interpretation of the observed amplitudes has to take into account the geometry of the stars and their modes. The strongly multiperiodic nature of \gdor\ pulsators complicates things further, since different modes in the same pulsator appear with different relative amplitudes depending on our viewing angle, that is, on the inclination of the rotation axis. Considering this, and the modest number of our spectroscopic sample targets, we only attempt to provide a qualitative interpretation below.

The dominant frequency, $f_1$, provides important additional context over the \vsi\--amplitude relation. In our spectroscopic sample, $f_1$ is strongly correlated with \vsi\ \revb{(Fig.~\ref{fig:f1-vs-vsini})}, while the photometric data shows that the highest observed amplitudes become less common at higher dominant frequencies \revb{(Fig.~\ref{fig:amp_vs_f1})}. Taken together, these relations can naturally contribute to the observed lack of high-amplitude, high-\vsi\ objects. In other words, part of the amplitude--\vsi\ pattern may be connected to the fact that stars with larger projected rotational velocities tend to have higher observed dominant frequencies, where the observable amplitude distribution is already more restricted. 

\rev{The apparent lack of low-\vsi, high-$f_1$ stars (top left part of Fig.~\ref{fig:f1-vs-vsini}) may provide information about the observed mode content. A low \vsi\ value can correspond either to slow rotation or to a more rapidly rotating star seen at a lower inclination. However, the high-inertial-frame frequencies of non-zonal prograde modes include a rotational Doppler-shift contribution. Such non-zonal modes are strongly cancelled in disk-integrated photometry near pole-on viewing, so rapid, low-inclination rotators are not expected to contribute their high-frequency non-zonal modes to this part of the diagram. A close-pole-on target can instead preferentially show an axisymmetric mode as dominant one, whose inertial frequency is lower, as it does not contain the azimuthal Doppler-shift term. Rotation, mode geometry, and disk integration can therefore change which modes have detectable photometric amplitudes and which one becomes the dominant peak \citep{Saio2018KIC5608334}. Period-spacing studies show that the observed mode-family content of \gdor\ stars changes with rotation, with prograde sectoral patterns dominating the detected population and some zonal or retrograde patterns mainly appearing among slow rotators \citep{Li_2020_gmode_near-core-rot}.}

\rev{In the highest \vsi\ members of our sample, equatorial rotation velocity is necessarily high, and also high inclination, near-equator-on viewing angle is statistically the most likely in this subset. According to both \cite{VanReeth2016} and \cite{Li_2020_gmode_near-core-rot}, faster near-core rotation implies more dominant prograde sectoral dipole modes, which are best visible at near-equator-on alignment. Yet, we see the lowest photometric amplitudes among these \gdor\ stars, both in PTP and in $A_1$.}

\rev{\cite{Aerts2025} provide a dominant-frequency proxy for near-core rotation when the dominant mode is a prograde dipole mode. Using this relation, Gaia DR3 radius estimations \citep{Gaia2023_DR3}, and assuming rigid body rotation, which, according to \cite{Li_2020_gmode_near-core-rot}, is a good approximation within 5\,\% of surface-to-near-core-rotation angular velocity relation, we can calculate approximate surface rotation velocities for each of our targets under the assumption that each $f_1$ belongs to a prograde dipole mode. The continuous red line in Fig.~\ref{fig:f1-vs-vsini} illustrates the $v_\mathrm{eq}$--$f_1$ relation on the same velocity scale used for \vsi. This line is the result of fitting the individual object data points. Since this line is a good lower-right boundary approximation of the observed data, which is expected as the inclination in \vsi\ is unknown, we can establish that the majority of our dominant frequencies belong indeed to prograde dipole modes, in accordance with the findings of \cite{VanReeth2016} and \cite{Li_2020_gmode_near-core-rot}. This is however a general trend only, and not individual mode identification.}

\subsection{\rev{Modelling the Inclination- and Spin-parameter-dependent Mode Visibilities}}
\label{sect:visibility-modelling}

\rev{To test whether the observed low pulsation amplitudes at high rotational velocities have real physical origin, or it is only the result of geometric effects, we modelled the inclination-dependent visibilities of sectoral prograde dipole modes for a wide variety of spin parameters.}

\rev{In our visibility toy model, we combined a scalar-visibility kernel computed from GYRE Hough functions \citep{TownsendTeitler2013GYRE} with the TAR visibilities of \citet{Townsend2003AngularModes,Townsend2003LightVariations}. We checked several different surface-to-near-core-rotation relations in the $0.8 < \alpha=f_{\rm core}/f_{\rm surface} < 1.2$ range, but since this had only marginal effect on the following results, we accepted a fixed $\alpha=1.0$. We also adopted a fixed linear limb darkening of $u=0.4$ as an illustrative example. We calculated scalar visibility kernels for a broad spin parameter range of $0<q<30$.}

\begin{figure}[!th]
  \centering
  \includegraphics[width=0.98\columnwidth]{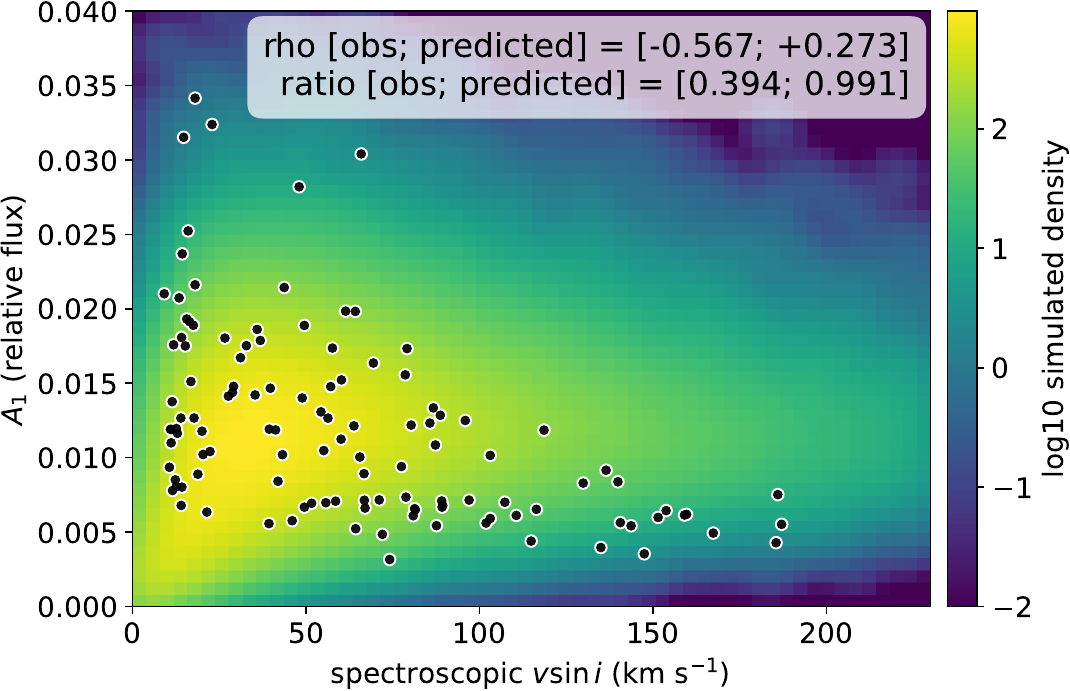}
  \caption{\rev{Visibility modelling of the amplitude of the dominant pulsation mode versus \vsi. The colour map represents the density distribution of the result of 2000 model runs (118\,000 individual simulated data points), while dots indicate the 119 spectroscopic observations with available radius estimate.}}
  \label{fig:sensitivity}
\end{figure}

\rev{Using this model, assuming a rotation-independent amplitude distribution, isotropic spatial inclination distribution, and a rotation velocity distribution that best describes our observed data in terms of amplitude and \vsi, we run Monte Carlo simulations of 2000 populations of 119 objects, based on the targets from our spectroscopic sample for which estimated radius were available in Gaia DR3. The obtained \vsi--$A_1$ model distribution, along with the actual observations is plotted in Fig.~\ref{fig:sensitivity}. The discrepancy is visible, but we also characterise the difference between the actual and the modelled distributions quantitatively.}

\rev{The Spearman rank correlation coefficient ($\rho$) between $A_1$ and \vsi\ for the observations is negative ($\rho_\mathrm{obs} = -0.567$), expressing anticorrelation, while definitely positive ($\rho_\mathrm{sim} = 0.273$) for the simulation. Thus, under the null hypothesis, amplitude is generally expected to increase with projected equatorial rotation velocity, while a definite decrease is observed in the data.}

\rev{The slope of the upper envelope is characterised by the ratio between the high (\vsi\,$>100$\,km s$^{-1}$) and low (\vsi\,$<50$\,km s$^{-1}$) q90 amplitude percentile. This ratio is 0.394 for the observations, expressing the visible decreasing trend, while 0.991 -- almost horizontal -- for the model prediction.}

\rev{We also performed a model compatibility check, calculating the $p$-value corresponding to the statistic-specific two-sided posterior-predictive-check, which show that the compatibility between the model and the observations is $p=0.001$, again, supporting the rejection of the null hypothesis.}

\rev{Note that our target selection imposed a low-amplitude cut-off, but the combined sample with that of \cite{Wang2025}, presented in the right-hand panel of Fig.~\ref{fig:wang-ace-context}, shows that the vast majority of the low-amplitude pulsators are below our threshold, and they can have the broadest range of \vsi. This fact confirms even stronger that the pure geometric model -- which trends towards an amplitude distribution maximum significantly higher than 0 at high \vsi\ -- is incompatible with the observations.}

\rev{In conclusion, the discrepancy between the observations and the visibility simulations rejects the null hypothesis. This means that the observed pulsation amplitude dependence on \vsi\ cannot be explained by geometric effects alone.}

\rev{Our observed $A_1$ vs. \vsi, $f_1$ vs. \vsi, $A_1$ vs. $f_1$, the earlier findings of \cite{VanReeth2016} and \cite{Li_2020_gmode_near-core-rot} and our visibility modelling together all indicate that the intrinsic, physical pulsation amplitudes of \gdor\ stars decrease with increasing equatorial rotational velocity.}

\subsection{\rev{HAGDOR Stars and Zonal Modes}}

\rev{\cite{Paunzen2020HAGDOR} investigated $T_{\rm eff}$ and $\log g$ of their initial HAGDOR sample on which the definition of the new subclass was founded. They did not find any obvious difference between HAGDORs and regular \gdor\ pulsators.}

\rev{Our present investigation adds an important nuance to this picture. We find that high observed pulsation amplitudes occur only in low-\vsi\ \gdor\ stars. Although this suggests correlation between HAGDOR nature and rotation, geometric effects once again complicate the picture. It cannot be excluded that some fast rotating \gdor\ stars can also be HAGDORs if viewed nearly pole-on, so that low inclination reduces \vsi.}

\begin{figure*}[!th]
  \centering
  \includegraphics[width=0.8\columnwidth]{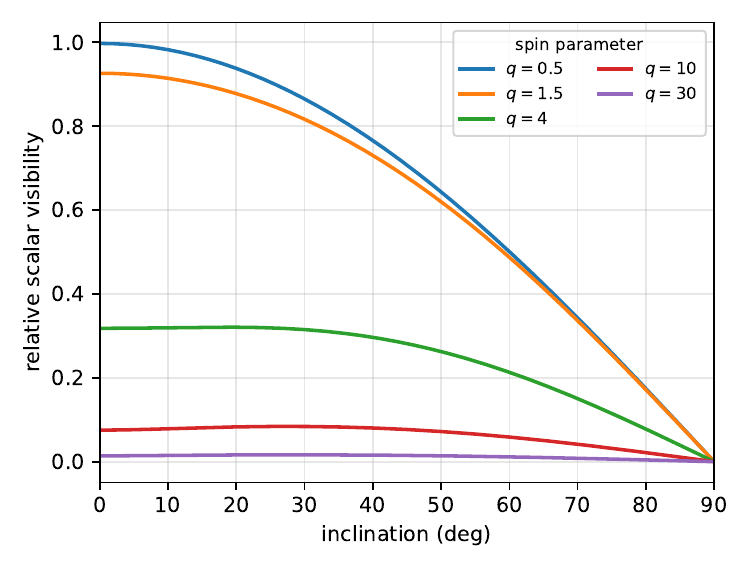}\hspace{20mm}
  \includegraphics[width=0.7\columnwidth]{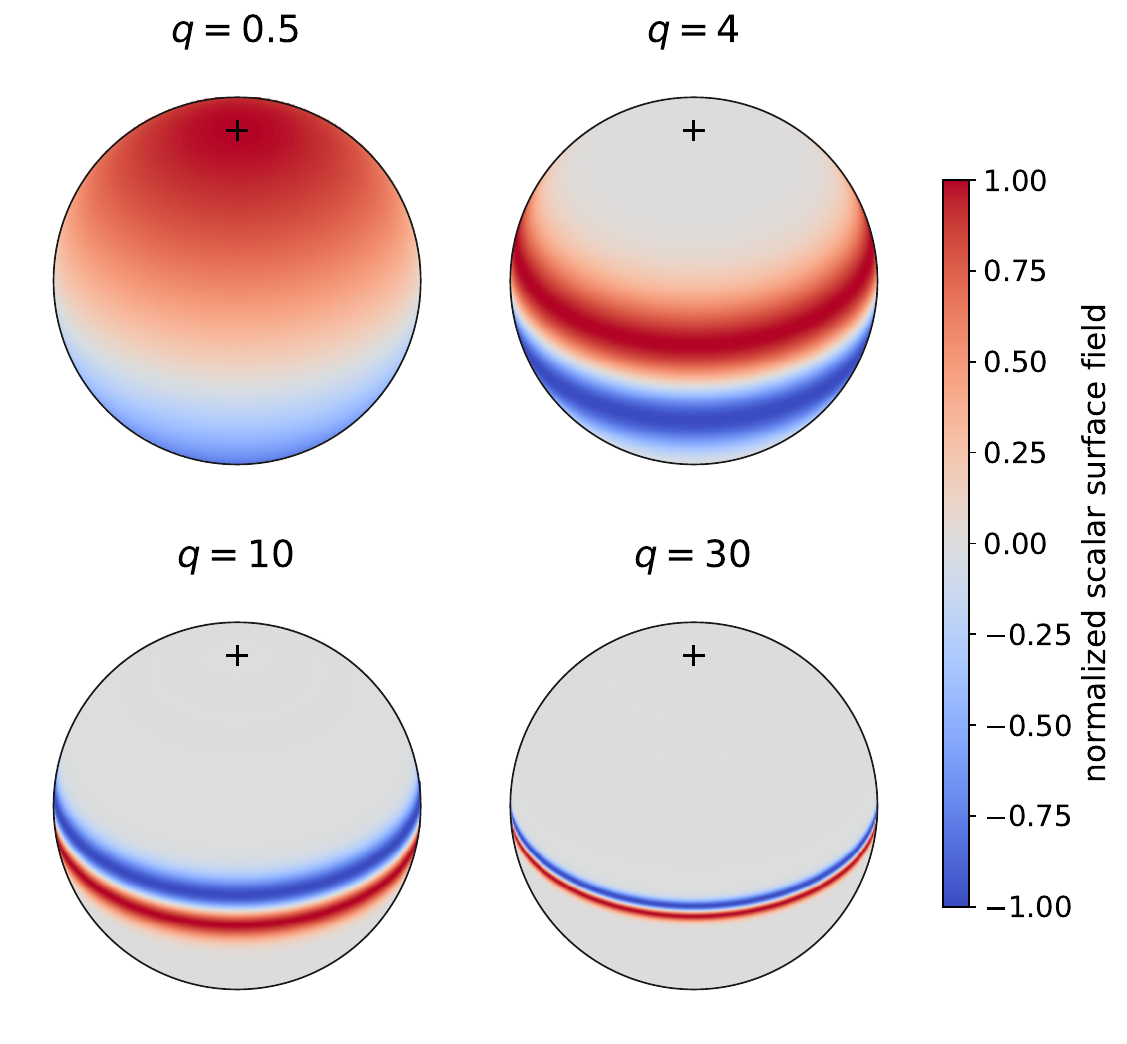}
  \caption{\rev{Modelling the dependence of zonal dipole mode visibilities on the spin parameter, $q$, and inclination, $i$. The left-hand panel shows the relative scalar visibilities normalized to a $q=0.5$ spin-parameter, $i=0$\,deg pole-on view. To the right, the visible stellar disk of four selected spin parameters are illustrated at $i=55$\,deg viewing angle. The scalar surface field is normalized so that the largest amplitude is unity.}}
  \label{fig:zonal-visibility}
\end{figure*}

\rev{To check this possibility, we also calculated relative visibilities for $\ell = 1, m = 0$ zonal dipole modes for different spin parameters, using the same assumptions listed in Sect.~\ref{sect:visibility-modelling} for sectoral dipole modes. The dependence of the mode visibility on inclination and spin parameter is plotted in Fig.~\ref{fig:zonal-visibility}.}

\rev{The calculations show that due to the phenomenon of equatorial trapping, meaning the increasing confinement of the pulsation scalar field to the vicinity of the equator with increasing spin parameter (see the right-hand panel of Fig.~\ref{fig:zonal-visibility}), the visibility of these modes strongly decreases at every viewing angle. Viewing from the poles, we see less and less of the high-amplitude regions of the scalar field, which get squeezed to the limb, while shifting the viewing angle towards the equator, the two hemispheres increasingly cancel each other out.}

\rev{The highest pulsation amplitudes are associated with low dominant pulsation frequencies. We calculated how the observed amplitude drops with increasing equatorial rotation velocity for two representative pulsation frequencies. We assumed $R\approx1.6\,R_\odot$ stellar radius and exact pole-on alignment of the viewing angle, since, except for the fastest rotators, this provides the best visibility of these modes. The chosen pulsation frequencies  were based on the HAGDOR sample of \cite{Paunzen2020HAGDOR}. Under the above assumptions, for a 0.5\,d$^{-1}$ and a 0.9\,d$^{-1}$ zonal dipole mode, the visibility of the observed amplitude drops to half of its maximal, non-rotating value at 60\,km\,s$^{-1}$ and 107\,km\,s$^{-1}$ equatorial velocity, respectively.}

\rev{In conclusion, the highest-amplitude \gdor\ stars are expected to be seen nearly pole on, and are slow or at most moderate rotators, where gravito-inertial effects are insignificant. Alternatively, if a HAGDOR is observed at not close to pole-on alignment, then it has to be a slow rotator.}

\rev{We find however, that HAGDORs are not quite a well-defined subclass of \gdor\ pulsators, as their defining boundary is arbitrary, in the sense that the transition from high- to lower-amplitude \gdor\ pulsations appears to be continuous, and the amplitude distribution does not indicate any bimodality. However, this is not in contradiction with the findings of \cite{Paunzen2020HAGDOR} on the phenomenological explanation of the highest-amplitude pulsations involving many superposed frequencies.}

\rev{We also note that Fig.~\ref{fig:zonal-visibility} demonstrates that the visibility of zonal dipole modes strongly decreases with increasing rotation velocity, regardless of the inclination, making them hardly detectable in fast rotating \gdor\ pulsators.}

\section{Summary}

We obtained \vsi\ measurements for 124 \gdor\ stars and combined them with Fourier parameters derived from TESS light curves.

\rev{We find that high observed pulsation amplitudes occur exclusively among stars with low \vsi, while the fastest rotators show only low photometric amplitudes. This amplitude--\vsi\ distribution is strongly linked to the observed frequency--rotation relation, where the dominant pulsation frequency increases with \vsi\ while the amplitude upper envelope decreases.}

\rev{Our visibility modelling indicates that these observed trends cannot be explained by geometric projection and mode visibility effects alone. The quantitative checks suggest that the intrinsic pulsation amplitudes of \gdor\ stars physically decrease with increasing equatorial rotational velocity. Furthermore, our analysis of zonal modes suggests that high-amplitude \gdor\ stars (HAGDORs) are likely either slow rotators or moderate rotators viewed nearly pole-on, as increasing rotation strongly suppresses the visibility of zonal dipole modes at any viewing angle.}

\textbf{Data availability.} The full machine-readable spectroscopic target catalogue and radial-velocity observations previewed in Tables~\ref{tab:spec124catalogue-sample} and \ref{tab:obslog-sample} will be available through the CDS/VizieR service \citep{Ochsenbein2000Vizier} upon publication. Public TESS and Gaia products used for cross-matching are available through their respective archive services. \rev{The supporting material, including the Wang reclassification tables and periodograms, and the scalar-visibility modelling package, will be available online at the ARP Data Repository upon publication (\url{https://hdl.handle.net/21.15109/ARP/ZW1ZLV}). The visibility-modelling package includes the code, copied inputs and outputs, parameter documentation, numerical-validation records, and reproduction instructions.}

\textbf{Software.} \rev{The calculations and supporting-material products use Python along with NumPy, SciPy, Matplotlib, Astropy, Astroquery, pandas, and h5py packages \citep{PythonSoftwareFoundation, Harris2020NumPy, Virtanen2020SciPy, Hunter2007Matplotlib, Astropy2022, Ginsburg2019Astroquery, McKinney2010pandas, Collette2013h5py}. The supporting periodograms use the Astropy Lomb--Scargle implementation, for which methodological context is given by \citet{VanderPlas2018LombScargle}.}

\begin{acknowledgements}
      This work has been supported by the KKP-137523 `SeismoLab' \'Elvonal grant of the Hungarian Research, Development and Innovation Office (NKFIH) and by the Lendület grant LP2012-31 of the Hungarian Academy of Sciences.
      This paper includes data collected by the TESS mission. Funding for the TESS mission is provided by NASA's Science Mission Directorate.
      This work has made use of data from the European Space Agency (ESA) mission Gaia (\url{https://www.cosmos.esa.int/gaia}), processed by the Gaia Data Processing and Analysis Consortium (DPAC, esa.int). Funding for the DPAC has been provided by national institutions, in particular the institutions participating in the Gaia Multilateral Agreement.
\end{acknowledgements}

\bibliographystyle{aa}
\bibliography{bibliography}

\end{document}